\documentclass[english]{article}
\usepackage[T1]{fontenc}
\usepackage[latin9]{inputenc}

\makeatletter

\providecommand{\tabularnewline}{\\}

\newcommand{\lyxaddress}[1]{
\par {\raggedright #1
\vspace{1.4em}
\noindent\par}
}

\usepackage{babel}
\makeatother

\begin{document}

\title{Neutrino Oscillation Probability from Tri-Bimaximality due to  Planck
Scale Effects}

\author{Bipin Singh Koranga }

\maketitle

\lyxaddress{Department of Physics, Kirori Mal college (University of Delhi,)
Delhi-110007, India}

\begin{abstract}
Current neutrino experimental data on neutrino mixing are well describes
by Tri-bi-maximal mixing, which is predicts $sin^{2}\theta_{12}=1/3,$zero
$U_{e3}$ and $\theta_{23}=45^{o}$. We consider the planck scale
operator on neutrino mixing. We assume that the neutrino masses and
mixing arise through physics at a scale intrimediate between planck
scale and the electroweak braking scale. We also assume, that just
above the electroweak breaking scale neutrino mass are nearly degenerate
and the mixing is tri-bi-maximal. Quantum gravity (Planck scale) effects
lead to an effective $SU(2)_{L}\times U(1)$ invariant dimension-5
Lagrangian symmetry involving Standard Model. On electroweak symmetry
breaking, this operator gives rise to correction to the neutrino masses
and mixings these additional terms can be considered as perturbation
to the tri-bimaximal neutrino mass matrix. We compute the deviation
of the three mixing angles and oscillation probability. We find that
the only large change in solar mixing angle and \% change in maximum
$P_{\mu e}$ is about 10\%.

\end{abstract}

\section{Introduction}

Recent advance in neutrino physics observation mainly of astrophysical
observation suggested the existence of tiny neutrino mass. The experiments
and observation has shown evidences for neutrino oscillation. The
solar neutrino deficit has been observed {[}1,2,3,4], the atmospheric
neutrino anomaly has been found {[}5,6,7], and currently almost confirmed
by KamLAND {[}8], and hence indicate that neutrino massive and there
is mixing in lepton sector, this indicate to imagine that there occurs
CP violation in lepton sector. Several physicist has considered whether
we can see CP violation effect in lepton sector through long baseline
neutrino oscillation experiments. The neutrino oscillation probabilities
in general depend on six parameters two independent mass squared difference
$\Delta_{21}$and $\Delta_{31}$, there mixing angle, $\theta_{12}$,
$\theta_{13}$., $\theta_{23}$.and one CP violating phase $\delta$.
There are two large mixing angle $\theta_{12}$, $\theta_{23}$and
one small ($\theta_{13}),$and two mass square difference $\Delta_{ij}=m_{j}^{2}-m_{i}^{2},$with
$m_{ij}$the neutrino masses.

Where

\begin{equation}
\Delta_{21}=\Delta_{solar},\end{equation}

\begin{equation}
\Delta_{31}=\Delta_{atm}.\end{equation}

The angle $\theta_{12}$ and $\theta_{23}$~represent the neutrino
mixing angles corresponding to solar and atmospheric neutrino oscillation.
Much progress has been made towards determining the values of the
three mixing angle. In this paper we discuss the effect of plancks
scale on neutrino mixing and neutrino oscillation probability.

\section{Neutrino Mixing Angle and Mass Squared Differences due to Planck
Scale Effects}

To calculate the effects of perturbation on neutrino observables.
The calculation developed in an earlier paper {[}12]. A natural assumption
is that unperturbed ($0^{th}$ order mass matrix) $M$~is given by

\begin{equation}
\mathbf{M}=U^{*}diag(M_{i})U^{\dagger},\end{equation}

where, $U_{\alpha i}$ is the usual mixing matrix and $M_{i}$ , the
neutrino masses is generated by Grand unified theory. Most of the
parameter related to neutrino oscillation are known, the major expectation
is given by the mixing elements $U_{e3}.$ We adopt the usual parametrization.

\begin{equation}
\frac{|U_{e2}|}{|U_{e1}|}=tan\theta_{12},\end{equation}

\begin{equation}
\frac{|U_{\mu3}|}{|U_{\tau3}|}=tan\theta_{23},\end{equation}

\begin{equation}
|U_{e3}|=sin\theta_{13}.\end{equation}

In term of the above mixing angles, the mixing matrix is

\begin{equation}
U=diag(e^{if1},e^{if2},e^{if3})R(\theta_{23})\Delta R(\theta_{13})\Delta^{*}R(\theta_{12})diag(e^{ia1},e^{ia2},1).\end{equation}

The matrix $\Delta=diag(e^{\frac{1\delta}{2}},1,e^{\frac{-i\delta}{2}}$)
contains the Dirac phase. This leads to CP violation in neutrino oscillation
$a1$ and $a2$ are the so called Majoring phase, which effects the
neutrino less double beta decay. $f1,$ $f2$ and $f3$ are usually
absorbed as a part of the definition of the charge lepton field. Planck
scale effects will add other contribution to the mass matrix that
gives the new mixing matrix can be written as {[}12]

\[
U^{'}=U(1+i\delta\theta),\]

\begin{equation}
=\left(\begin{array}{ccc}
U_{e1} & U_{e2} & U_{e3}\\
U_{\mu1} & U_{\mu2} & U_{\mu3}\\
U_{\tau1} & U_{\tau2} & U_{\tau3}\end{array}\right)+i\left(\begin{array}{ccc}
U_{e2}\delta\theta_{12}^{*}+U_{e3}\delta\theta_{23,}^{*} & U_{e1}\delta\theta_{12}+U_{e3}\delta\theta_{23}^{*}, & U_{e1}\delta\theta_{13}+U_{e3}\delta\theta_{23}^{*}\\
U_{\mu2}\delta\theta_{12}^{*}+U_{\mu3}\delta\theta_{23,}^{*} & U_{\mu1}\delta\theta_{12}+U_{\mu3}\delta\theta_{23}^{*}, & U_{\mu1}\delta\theta_{13}+U_{\mu3}\delta\theta_{23}^{*}\\
U_{\tau2}\delta\theta_{12}^{*}+U_{\tau3}\delta\theta_{23}^{*}, & U_{\tau1}\delta\theta_{12}+U_{\tau3}\delta\theta_{23}^{*}, & U_{\tau1}\delta\theta_{13}+U_{\tau3}\delta\theta_{23}^{*}\end{array}\right).\end{equation}

Where $\delta\theta$ is a hermition matrix that is first order in
$\mu${[}12,13]. The first order mass square difference $\Delta M_{ij}^{2}=M_{i}^{2}-M_{j}^{2},$get
modified {[}12,13] as

\begin{equation}
\Delta M_{ij}^{'^{2}}=\Delta M_{ij}^{2}+2(M_{i}Re(m_{ii})-M_{j}Re(m_{jj}),\end{equation}

where

\[
m=\mu U^{t}\lambda U,\]

\[
\mu=\frac{v^{2}}{M_{pl}}=2.5\times10^{-6}eV.\]

The change in the elements of the mixing matrix, which we parametrized
by $\delta\theta${[}12], is given by

\begin{equation}
\delta\theta_{ij}=\frac{iRe(m_{jj})(M_{i}+M_{j})-Im(m_{jj})(M_{i}-M_{j})}{\Delta M_{ij}^{'^{2}}}.\end{equation}

The above equation determine only the off diagonal elements of matrix
$\delta\theta_{ij}$. The diagonal element of $\delta\theta_{ij}$
can be set to zero by phase invariance. Using Eq(8), we can calculate
neutrino mixing angle due to Planck scale effects,

\begin{equation}
\frac{|U_{e2}^{'}|}{|U_{e1}^{'}|}=tan\theta_{12}^{'},\end{equation}

\begin{equation}
\frac{|U_{\mu3}^{'}|}{|U_{\tau3}^{'}|}=tan\theta_{23}^{'},\end{equation}

\begin{equation}
|U_{e3}^{'}|=sin\theta._{13}^{'}\end{equation}

For degenerate neutrinos, $M_{3}-M_{1}\cong M_{3}-M_{2}\gg M_{2}-M_{1},$
because $\Delta_{31}\cong\Delta_{32}\gg\Delta_{21}.$ Thus, from the
above set of equations, we see that $U_{e1}^{'}$ and $U_{e2}^{'}$
are much larger than $U_{e3}^{'},\,\, U_{\mu3}^{'}$ and $U_{\tau3}^{'}$.
Hence we can expect much larger change in $\theta_{12}$ compared
to $\theta_{13}$ and $\theta_{23}.$ As one can see from the above
expression of mixing angle due to Planck scale effects, depends on
new contribution of mixing $U^{'}=U(1+i\delta\theta).$ We assume
that, just above the electroweak breaking scale, the neutrino masses
are nearly degenerate and the mixing are Tri-bimaximal, with the value
of the mixing angle as $\theta_{12}=35^{o},\,\,\theta_{23}=\pi/4$
and $\theta_{13}=0.$ Taking the common degenerate neutrino mass to
be 2 eV, which is the upper limit coming from tritium beta decay {[}9].
We compute the modified mixing angles using Eqs (11)-(13). We have
taken $\Delta_{31}=0.002eV^{2}[10]$ and $\Delta_{21}=0.00008eV^{2}${[}11].
For simplicity we have set the charge lepton phases $f_{1}=f_{2}=f_{3}=0.$~Since
we have set the $\theta_{13}=0,$ the Dirac phase $\delta$~drops
out of the zeroth order mixing angle. Next section , we discuss the
neutrino oscillation probability under Planck scale effects

\section{Neutrino Oscillation Probability Under Planck Scale Effects}

The flux of solar neutrino observed by the Homestake detector was
on third of that predicted by Standard solar Model (SSM). The phenomenon
of neutrino oscillation can be used to explain neutrino deficit. suppose
an electron neutrino is produced at $t=0.$ A set of neutrino mass
eigen state at $t=0$ as

\begin{equation}
|\nu(t=0)>|\nu_{e}>=cos\theta_{12}|\nu_{1}(0)>+sin\theta_{12}|\nu_{2}(0)>.\end{equation}

After time $t$ it becomes

\begin{equation}
|\nu(t=t)>|\nu_{\mu}>=cos\theta_{12}e^{-iE_{1}t}|\nu_{1}(0)>+sin\theta_{12}e^{-iE_{2}t}|\nu_{2}(0)>.\end{equation}

Then the oscillation probability becomes

\begin{equation}
P(\nu_{e}\rightarrow\nu_{\mu})=sin^{2}2\theta_{12}sin^{2}\left(\frac{1.27\Delta_{21}L}{E}\right),\end{equation}

and the survival probability

\begin{equation}
P(\nu_{e}\rightarrow\nu_{e})=1-sin^{2}2\theta_{12}sin^{2}\left(\frac{1.27\Delta_{21}L}{E}\right).\end{equation}

In the above two equation units of $\Delta_{21}=m_{2}^{2}-m_{1}^{2}$
is $ev^{2},$L (baseline length) is in meter and E is neutrino energy
in MeV. For a maximum oscillation case the phase term in eq(16), $\left(\frac{1.27\Delta_{21}L}{E}\right)$
equal to $\frac{\pi}{2}$, then oscillation probability only depend
on $\theta_{12}$ 

\begin{equation}
P(\nu_{e}\rightarrow\nu_{\mu})=sin^{2}2\theta_{12}.\end{equation}

The oscillation probability due to Planck scale effects is

\begin{equation}
P(\nu_{e}\rightarrow\nu_{\mu})=sin^{2}2\theta_{12}^{'},\end{equation}

In the above Eq(19), $\theta_{12}^{'}$ is the mixing angle due to
Planck scale effects.

\section{Numerical Results}

We assume that, just above the electroweak breaking scale, the neutrino
masses are nearly degenerate and the mixing are Teri-bi maximal, with
the value of the mixing angle as $\theta_{12}=35^{o},\,\theta_{23}=\pi/4$
and $\theta_{13}=0.$ Taking the common degenerate neutrino mass to
be 2 eV, which is the upper limit coming from tritium beta decay {[}9].
We compute the modified mixing angles using Eqs (11)-(13). We have
taken $\Delta_{31}=0.002eV^{2}[10]$ and $\Delta_{21}=0.00008eV^{2}${[}11].
For simplicity we have set the charge lepton phases $f_{1}=f_{2}=f_{3}=0.$~Since
we have set the $\theta_{13}=0,$ the Dirac phase $\delta$~drops
out of the zeroth order mixing angle. We compute the modified mixing
angles as function of $a_{1}$ and $a_{2}$ using Eq(11). In table
1, we list the modified neutrino mixing angle $\theta_{12}^{'}$ and
maximum $P(\nu_{e}\rightarrow\nu_{\mu})$ oscillation probability
for some sample of $a_{1}$ and $a_{2}$.

\begin{table}
\begin{tabular}{|c|c|c|c|}
\hline 
$a_{1}$ & $a_{2}$ & $\theta_{12}^{'}$ & $P(\nu_{e}\rightarrow\nu_{\mu})=sin^{2}2\theta_{12}^{'}$\tabularnewline
\hline
\hline 
$0^{o}$ & $0^{o}$ & $36.63^{o}$ & $0.94$\tabularnewline
\hline 
$0^{o}$ & $45^{o}$ & $36.82^{o}$ & $0.92$\tabularnewline
\hline 
$0^{o}$ & $90^{o}$ & $34.99^{o}$ & $0.88$\tabularnewline
\hline 
$0^{o}$ & $135^{o}$ & $36.88^{o}$ & $0.87$\tabularnewline
\hline 
$0^{o}$ & $180^{o}$ & $38.51^{o}$ & $0.94$\tabularnewline
\hline 
$45^{o}$ & $0^{o}$ & $36.63^{o}$ & $0.91$\tabularnewline
\hline 
$45^{o}$ & $45^{o}$ & $34.97^{o}$ & $0.96$\tabularnewline
\hline 
$45^{o}$ & $90^{o}$ & $33.26^{o}$ & $0.84$\tabularnewline
\hline 
$45^{o}$ & $135^{o}$ & $35.09^{o}$ & $0.88$\tabularnewline
\hline 
$45^{o}$ & $180^{o}$ & $36.63^{o}$ & $0.91$\tabularnewline
\hline 
$90^{o}$ & $0^{o}$ & $35^{o}$ & $0.88$\tabularnewline
\hline 
$90^{o}$ & $45^{o}$ & $33.43^{o}$ & $0.84$\tabularnewline
\hline 
$90^{o}$ & $90^{o}$ & $31.77^{o}$ & $0.80$\tabularnewline
\hline 
$90^{o}$ & $135^{o}$ & $33.49^{o}$ & $0.84$\tabularnewline
\hline 
$90^{o}$ & $180^{o}$ & $35^{o}$ & $0.88$\tabularnewline
\hline 
$135^{o}$ & $0^{o}$ & $36.63^{o}$ & $0.91$\tabularnewline
\hline 
$135^{o}$ & $45^{o}$ & $35.04^{o}$ & $0.89$\tabularnewline
\hline 
$135^{o}$ & $90^{o}$ & $33.26^{o}$ & $0.84$\tabularnewline
\hline 
$135^{o}$ & $135^{o}$ & $35.02^{o}$ & $0.88$\tabularnewline
\hline 
$135^{o}$ & $180^{o}$ & $36.63^{o}$ & $0.91$\tabularnewline
\hline 
$180^{o}$ & $0^{o}$ & $38.51^{o}$ & $0.94$\tabularnewline
\hline 
$180^{o}$ & $45^{o}$ & $36.82^{o}$ & $0.92$\tabularnewline
\hline 
$180^{o}$ & $90^{o}$ & $34.99^{o}$ & $0.92$\tabularnewline
\hline 
$180^{o}$ & $135^{o}$ & $36.88^{o}$ & $0.87$\tabularnewline
\hline 
$180^{o}$ & $180^{o}$ & $38.51^{o}$ & $0.94$\tabularnewline
\hline
\end{tabular}

\caption{Modified mixing angles and maximum $P(\nu_{e}\rightarrow\nu_{\mu})$
oscillation probabilities for some sample of $a_{1}$ and $a_{2}$.
Input value are $\Delta_{31}=0.002eV^{2},$ $\Delta_{21}=0.00008eV^{2},$$\theta_{12}=35^{o},\theta_{23}=45^{o},$$\theta_{13}=0^{o}.$}

\end{table}

From Table 1, we see that planck scale effects change the $\theta_{12}$~from
the Tri-bimaximal value of $\theta_{12}=35^{o}$to a value close the
the best fit value of the data {[}15,16]. The Planck scale effects
give rise the correction to neutrino mass matrix on electroweak symmetry
breaking. It is imperative to cheack that these correction do not
spoil the good agreement between the experiments fits and the predection
of the tri-bimaximal mixing scenorio. It is expected that dynamics
at a higher scale generates the neutrino mass matrix, which will eventually
provides the presently observed neutrino mass and mixing. In an attractive
scenario, the neutrino mixing pattern generated by high scale dynamics
is predicted to be tri-bimaximal. However the solar neutrino data
show that the mixing angle $\theta_{12}$ is substantially less than
$35^{o}$. It is argued in the literature that renormalization group
evolution effects from the higher scale to electroweak scale, can
bring down the value of $\theta_{12}$ from $35^{o}$ to a value which
is within the experimentally acceptable range. However, for a large
range of neutrino parameters, the renormalization group evolution
leads to negligible change in the neutrino mass matrix. Then it become
imperative to search for such alternate mechanism for which the necessary
reduction in $\theta_{12}$ can be achieved.

\section{Conclusions}

In this paper, we studied, how Planck scale effects the mixing and
oscillation probability. The effective dimension-5 operator from Planck
scale {[}12], leads to correction in neutrino mass matrix at the electroweak
symmetry breaking scale. We compute the change in the mixing angle
due to additional mass terms for the case of Tri-bimaximal. The change
in $\theta_{12}$ is more than $3^{o}$from the Tri-bimaximal value.
Therefore corresponding maximum change in oscillation probability
is about 10\%. The change of $\theta_{12}$occurs of course, for degenerate
neutrino mass with a common mass of about 2 eV. Cosmology constraints,
from WMAP measurement {[}14] impose an upper limit of 0.7eV on neutrino
mass.Then the change in the value of $\theta_{12}$is smaller. One
summarizing statement of this work might be the following, due to
Planck scale effects only $\theta_{12}$ deviated by $3.5^{o}$ and
other mixing angle have very small deviation and maximum change of
$P(\nu_{e}\rightarrow\nu_{\mu})$ oscillation probability is about
10\%, this can be achieved by our calculation of {}``Tri-Bimaximal''
neutrino mixing.

\end{document}